\documentclass[floatfix,groupedaddress,superscriptaddress,aps,showpacs,amsmath,amssymb,floatfix, prl, longbibliography, twocolumn,a4]{revtex4-2}

\usepackage{graphicx,float}
\usepackage{subfigure}
\usepackage{dcolumn}
\usepackage{bm}
\usepackage{mathrsfs}
\usepackage{txfonts}
\usepackage{CJK}
\usepackage[amsmath]{SIunits}
\usepackage{epsfig}
\usepackage{epstopdf}
\usepackage{color}
\usepackage{placeins}
  \usepackage{lineno}
\usepackage{lipsum}

\usepackage[colorlinks,
linkcolor=blue,
anchorcolor=blue,
citecolor=blue,
filecolor=blue,
menucolor=blue,
runcolor=blue,
urlcolor=blue,
frenchlinks=blue]{hyperref}

\allowdisplaybreaks[2]
\newenvironment{SChinese}{%
\CJKfamily{gbsn}%
\CJKtilde
\CJKnospace}{}
\allowdisplaybreaks[2]

\begin{document}

\begin{CJK}{UTF8}{}
\begin{SChinese}

\title{Quantum Cross Nonlinearity for Photon-Number-Resolving Nondestructive Detection}

\author{Jiang-Shan Tang}  %
 \affiliation{College of Engineering and Applied Sciences, National Laboratory of Solid State Microstructures, and Collaborative Innovation Center of Advanced Microstructures, Nanjing University, Nanjing 210023, China}
  \affiliation{Hefei National Laboratory, Hefei 230088, China}

\author{Mingyuan Chen}  %
 \affiliation{College of Engineering and Applied Sciences, National Laboratory of Solid State Microstructures, and Collaborative Innovation Center of Advanced Microstructures, Nanjing University, Nanjing 210023, China}

\author{Miao Cai}  %
 \affiliation{College of Engineering and Applied Sciences, National Laboratory of Solid State Microstructures, and Collaborative Innovation Center of Advanced Microstructures, Nanjing University, Nanjing 210023, China}

\author{Lei Tang}  %
 \affiliation{College of Engineering and Applied Sciences, National Laboratory of Solid State Microstructures, and Collaborative Innovation Center of Advanced Microstructures, Nanjing University, Nanjing 210023, China}

\author{Yan-Qing Lu}  %
\email{yqlu@nju.edu.cn}
     \affiliation{College of Engineering and Applied Sciences, National Laboratory of Solid State Microstructures, and Collaborative Innovation Center of Advanced Microstructures, Nanjing University, Nanjing 210023, China}

\author{Keyu Xia}  %
 \email{keyu.xia@nju.edu.cn}
    \affiliation{College of Engineering and Applied Sciences, National Laboratory of Solid State Microstructures, and Collaborative Innovation Center of Advanced Microstructures, Nanjing University, Nanjing 210023, China}
  \affiliation{Hefei National Laboratory, Hefei 230088, China}
  \affiliation{Shishan Laboratory, Suzhou Campus of Nanjing University, Suzhou 215000, China}

\author{Franco Nori}
\email{fnori@riken.jp}
 \affiliation{Quantum Computing Center, Cluster for Pioneering Research, RIKEN, Wako-shi, Saitama 351-0198, Japan}
 \affiliation{Physics Department, The University of Michigan, Ann Arbor, Michigan 48109-1040, USA}

\date{\today}

\begin{abstract}
We present an unconventional mechanism for quantum nonlinearity in a system comprising of a V-type quantum emitter (QE) and two Fabry-P\'{e}rot cavities. The two transitions of the V-type QE are effectively coupled with two independent cavity modes. The system exhibits a strong quantum nonlinear control in the transmission even at the single-photon level, which we refer to as \emph{quantum cross nonlinearity}. The underlying physics can be understood as \emph{quantum competition} between the two transitions of the QE sharing a common ground state. By leveraging this quantum cross nonlinearity, we further show photon-number-resolving quantum nondestructive detection.
Owing to the widespread nature of this V-type configuration, our approach can be readily extended to diverse cavity quantum electrodynamic systems beyond the realm of optics, encompassing, e.g., microwave photons and acoustic wave phonons. This versatility may facilitate numerous unique applications for quantum information processing.
\end{abstract}

\maketitle

\end{SChinese}
\end{CJK}

\emph{Introduction}.---Achieving strong nonlinearity between two electromagentic fields, particularly at the single-photon level, is a longstanding goal in quantum optics, but remains a challenge~\cite{nature.488.57.2012}. Attempts to access a strong nonlinearity in a classical optical medium have been proven exceedingly challenging for light fields with a small number of photons~\cite{natphotons.8.685.2014}.

Relevant advancements in quantum optics have illuminated a viable pathway to attain nonlinearity at the quantum level. The exploitation of anharmonic energy-level splitting, mechanical motion and the excitation-saturation effect in two-level quantum emitters (QEs) to induce quantum nonlinearity has garnered widespread attention, because of the important applications in photon blockade and quantum nonreciprocity~\cite{nature.436.87.2005, science.319.1062.2008, PhysRevLett.107.063601.2011, PhysRevLett.113.243601.2014, PhysRevA.92.033806.2015, NatPyhs.14.885.2018, PhysRevLett.121.153601.2018, nature.569.2019, PhysRevLett.123.233604.2019, lpr.16.2100430.2022, PhysRevLett.128.203602.2022, natphysLiu2024}. Giant Kerr nonlinearity based on electromagnetically induced transparency has also been well understood through quantum interference effects~\cite{PhysRevA.51.576.1995, OptLett.21.1936, Phys.Rev.Lett.79.1467.1997, PhysRevA.65.063804.2002, PhysRevLett.91.093601, PhysRevLett.100.173602.2008, PhysRevLett.103.150503.2009, science.333.1266.2011, natcommun.5.5082.2014, PhysRevA.95.063849.2017, PhysRevApplied.15.064020.2021}, but mostly accessible in an ensemble of N-type atoms except for superconducting artificial atoms~\cite{PhysRevLett.103.150503.2009}.
Therefore, the demand for novel quantum nonlinear mechanisms remains a longstanding and ongoing endeavor.
These efforts hold great promise for both fundamental research and technological applications, including single-photon switches and transistors~\cite{nature.460.76.2009,natphotons.6.605.2012}, all-optical deterministic quantum logic gates~\cite{Light.1.e40.2012,nature.508.237.2014}, and strongly correlated states~\cite{nature.502.71.2013,natphys.16.921.2020,PhysRevLett.125.143605.2020}.
Additionally, these methods provide an opportunity for mutual control of quantum probe fields~\cite{science.320.769.2008,PhysRevLett.128.083604.2022}, crucial for quantum nondemolition (QND) measurement~\cite{nature.400.239.1999}.

Photon-number-resolving measurements are essential for studying fundamental quantum physics and quantum information technologies. However, most of these measurements rely on the absorption of photons~\cite{natphotonics.17.106.2023,natphotonics.17.112.2023,PhysRevLett.132.203601.2024}. QND measurement of photons promises more exotic applications as it allows for the repeated use of photons, thereby greatly enhancing quantum information processing.
For an electromagnetic-field mode, nondestructive detection involves monitoring the photon number of signal fields without changing them~\cite{PhysRevLett.65.976.1990,scullyquantum.1997,agarwalquantum.2012,science.342.1349.2013}. Thanks to the enhanced light-matter interaction within cavities, QND detection of photons has been experimentally demonstrated in various configurations in the microwave domain, including dispersion-shifted two-level QEs~\cite{natphys.14.546.2018} and phase-sensitive ladder three-level QEs~\cite{nature.400.239.1999,science.342.1349.2013,PhysRevLett.112.093601.2014}. The application of optical Kerr nonlinearity to QND measurement is still under debate~\cite{PhysRevA.73.062305.2006,PhysRevA.81.043823.2010}, but single-photon QND can be achieved by combining the quantum nonlocal response with coherently amplified optical Rabi-like coupling in a $\chi^{(3)}$ nonlinear medium~\cite{PhysRevLett.116.023601.2016,PhysRevA.97.032314.2018,PhysRevLett.125.243601.2020}. These methods ingeniously harness the nonlinear response encoded in the phase of the probe fields~\cite{nature.448.889.2007}. In stark contrast, the accompanying intensity information of the probe fields is equally important but exclusive in the context of QND measurements. Hence, it is highly desirable to reveal new mechanisms for nondestructive detection of photons by exploring the quantum nonlinearity directly encoded in the probe intensity.

V-type QEs have been explored for achieving QND readout of the qubit state based on the shelving effect ~\cite{PhysRevLett.100.200502.2008,Natcommun.12.6383.2021}. Here, \emph{without} the shelving effect, we show \emph{quantum cross nonlinearity} (QCN) induced in a V-type system, manifested by a mutual nonlinear modulation of the transmission of two probe fields. This quantum nonlinearity originates from the quantum competition between two transition channels sharing the common ground state. QCN is further utilized for photon-number-resolving QND measurements. For a signal field with a small number of photons, the survival rate can exceed $92\%$ by using existing experimental technologies.

\begin{figure}
	\centering
	\includegraphics[width=1.0\linewidth]{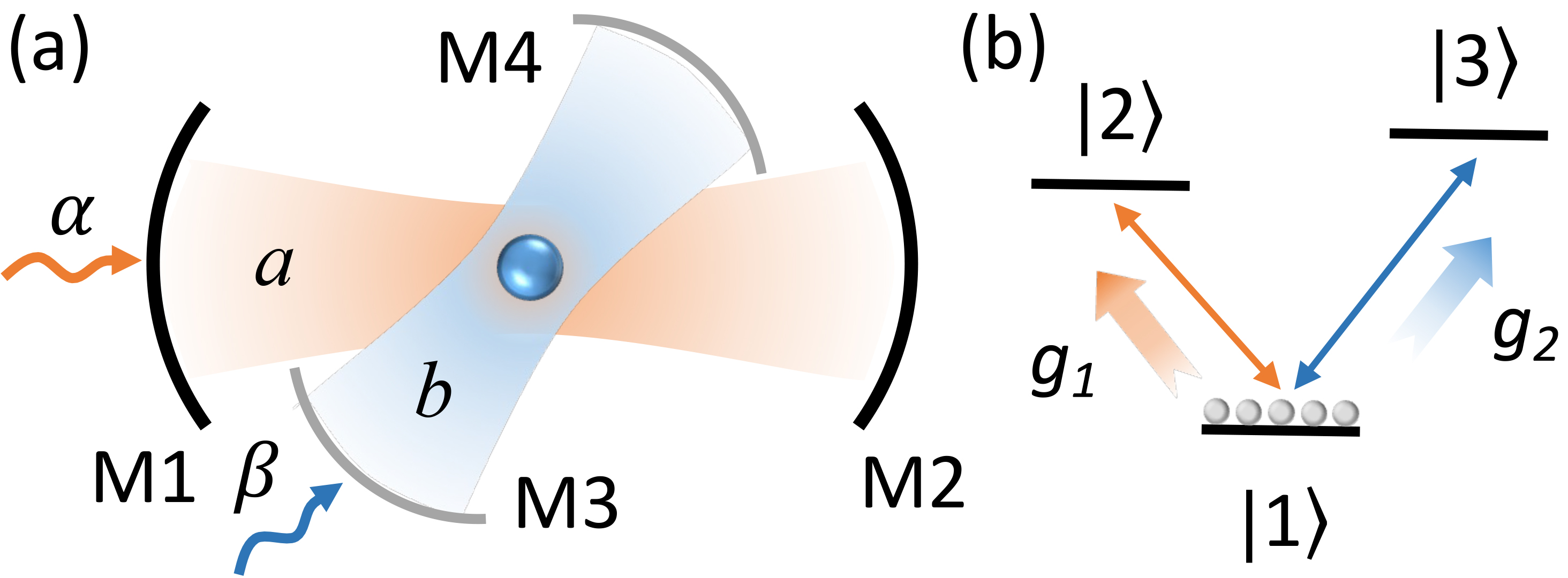} \\
	\caption{(a) Schematic diagram for quantum cross nonlinearity and QND measurement of photons. The system consists of two Fabry-P\'{e}rot (F-P) cavities, comprising four mirrors M1--M4, and simultaneously coupled to a V-type quantum emitter (QE). The cavity modes are excited by two probe fields $\alpha$ and $\beta$, respectively. (b) Level diagram of the V-type QE, of which two transitions couple to two cavity modes with strengths $g_1$ and $g_2$, respectively.}
	\label{fig:FIG1}
\end{figure}
\emph{System and model}.---The system configuration is schematically shown in Fig.~\ref{fig:FIG1}. The two independent Fabry-P\'{e}rot (F-P) cavities are formed by two pairs ( M1--M2, and M3--M4) of highly reflective mirrors, causing an external loss denoted by $\kappa_{\text{ex},i}$ ($i=1, 2, 3, 4$). The two cavity modes, denoted as $a$ and $b$, are characterized by their resonant frequencies, $\omega_a$ and $\omega_b$, respectively. The two cavities simultaneously couple with a V-type QE. This QE comprises of a ground state $|1\rangle$ and two excited states $|2\rangle$ and $|3\rangle$, with corresponding eigenfrequencies $\omega_1$, $\omega_2$, and $\omega_3$, forming two distinct transitions: $|1\rangle\leftrightarrow|2\rangle$ and $|1\rangle\leftrightarrow|3\rangle$, as depicted in Fig.~\ref{fig:FIG1}(b). The cavity modes $a$ and $b$ couple to distinct transitions of the QE with coupling strengths of $g_1$ and $g_2$, respectively. Such V-type QEs can be realized with atoms~\cite{PhysRevA.90.043802.2014,PhysRevX.5.041036.2015,science.354.1577.2016,PhysRevA.99.043833.2019,lpr.16.2100708.2022}, ions~\cite{PhysRevLett.100.200502.2008}, and artificial atoms~\cite{PhysRevB.75.104516.2007,science.317.929.2007,PhysRevA.89.063822.2014,nature.569.2019,natmater.18.1065.2019,Natcommun.12.6383.2021}, ensuring the broad applicability of our proposal.

We now consider two weak coherent inputs, $\alpha$ and $\beta$, with frequencies $\omega_{p_1}$ and $\omega_{p_2}$. These fields excite cavity modes $a$ and $b$ through mirrors M1 and M3, respectively.
We denote $\Delta_{a}= (\omega_2-\omega_1) -\omega_{a}$ and $\Delta_{b}= (\omega_3-\omega_1) -\omega_{b}$ as the detunings for the cavity modes $a$ and $b$ with respect to the transitions $|1\rangle\leftrightarrow|2\rangle$ and $|1\rangle\leftrightarrow|3\rangle$, respectively. The detunings between the cavity modes and the drivings are $\Delta_{1}=\omega_a-\omega_{p_1}$ and $\Delta_{2}=\omega_b-\omega_{p_2}$, yielding $\Delta_{3}=\Delta_{a}+\Delta_{1}$, and $\Delta_{4}=\Delta_{b}+\Delta_{2}$.
In the rotating frame, defined by the unitary transformation $U=\text{exp}\{i[\omega_{p_1}a^{\dagger}a+\omega_{p_2}b^{\dagger}b+\omega_1\sigma_{11}+(\omega_1+\omega_{p_1})\sigma_{22}
+(\omega_1+\omega_{p_2})\sigma_{33}]t\}$, the system Hamiltonian takes the form ($\hbar=1$) $H_\text{S} = H_\text{QCN} + H_\text{D}$, with the QE-cavity part $H_\text{QCN} = \Delta_1a^{\dagger}a+\Delta_2 b^{\dagger}b+\Delta_3\sigma_{22}+\Delta_4\sigma_{33} + g_1\left(\sigma_{21}a+a^{\dagger}\sigma_{12}\right) +g_2\left(\sigma_{31}b+b^\dagger\sigma_{13}\right)$, for inducing the QCN, and the driving part $H_\text{D} =  i\sqrt{\kappa_{\text{ex}, 1}}\left(\alpha^{\ast}a-\alpha a^\dagger\right)+i\sqrt{\kappa_{\text{ex}, 3}}\left(\beta^{\ast}b-\beta b^\dagger\right)$, for probing. Here, $\sigma_{mn}=|m\rangle\langle n|$, $m$ and $n \in \{1, 2, 3\}$.
The system dynamics can be described by the quantum master equation of the density matrix $\rho$ as
\begin{equation}\label{eq:density}
\dot{\rho}=-i\left[H_\text{S}, \rho\right]+ \mathcal{L}\left\{\Gamma, O\right\}\rho\;,
\end{equation}
where $\Gamma=\left\{\kappa_a, \kappa_b, \gamma_{21}, \gamma_{31}\right\}$ and $O=\left\{a, b, \sigma_{12}, \sigma_{13}\right\}$, $\gamma_{21}$ and $\gamma_{31}$ describe the decay rates of the two excited states to the ground state, respectively. The Lindblad operator is expressed as $\mathcal{L}\left\{\Gamma, O\right\}\rho=\left(\Gamma/2\right)\left(2O\rho O^\dagger-O^\dagger O\rho-\rho O^\dagger O\right)$. The total losses of the cavity modes $a$ and $b$ are respectively given by $\kappa_a=\kappa_{\text{ex}, 1}+\kappa_{\text{ex}, 2}+\kappa_{\text{in}, a}$ and $\kappa_b=\kappa_{\text{ex}, 3}+\kappa_{\text{ex}, 4}+\kappa_{\text{in}, b}$. The intrinsic losses, $\kappa_{\text{in}, a}$ and $\kappa_{\text{in}, b}$  of cavity fields, are negligible.

Using the input-output relations $a_t=\sqrt{\kappa_{\text{ex},2}}a$ and $b_t=\sqrt{\kappa_{\text{ex},4}}b$~\cite{PhysRevA.31.3761.1985,PhysRevLett.70.2269.1993}, we can derive the analytical expressions for the steady-state transmissions, denoted as $T_a = \langle a_{t}^{\dagger}a_t\rangle/|\alpha|^{2}$ and $T_b = \langle b_{t}^{\dagger}b_t\rangle/|\beta|^{2}$~\cite{SupplMat}, respectively. Furthermore, a fully quantum treatment allows us to numerically solve Eq.~\eqref{eq:density} by truncating the Fock states of the cavity fields~\cite{science.319.1062.2008}.

Next, we examine the time-dependent dynamic evolution of the system when the input signal is a quantum field with the quantum cascaded method~\cite{PhysRevLett.70.2273.1993,PhysRevLett.70.2269.1993,Carmichaelquantum.1999,PhysRevA.106.063707.2022,SupplMat}. In simulations, the coherent input fields are modeled by fields from two virtual ``source'' cavities $d_1$ and $d_2$, which can numerically generate quantum fields. We use the Hamiltonians $H_{d_1}$ and $H_{d_2}$ (see the supplementary material) modeling the source cavities $d_1$ and $d_2$ to replace the classical driving $H_{\text{D}}$. The master equation describing the quantum cascaded system reads
\begin{equation}\label{eq:cascadedmaster}
 \begin{split}
  \dot{\rho}_{\text{qcs}} &= -i\left[H_{\text{QCN}}, \rho_{\text{qcs}}\right]-i\left[H_{d_1}, \rho_{\text{qcs}}\right]-i\left[H_{d_2}, \rho_{\text{qcs}}\right] \\
   &+\mathcal{L}\left\{\Gamma_{\text{qcs}}, O_{\text{qcs}}\right\}\rho_{\text{qcs}}+\mathcal{L}_{\text{Net},a}\rho_{\text{qcs}}+\mathcal{L}_{\text{Net},b}\rho_{\text{qcs}}\;,
 \end{split}
\end{equation}
where the symbol $\rho_{\text{qcs}}$ is the joint density matrix of the source cavity modes and the main system for the QCN, the Lindblad terms are $\mathcal{L}\left\{\Gamma_{\text{qcs}}, O_{\text{qcs}}\right\}\rho_{\text{qcs}}=(\Gamma_{\text{qcs}}/2)\left(2O_{\text{qcs}}\rho_{\text{qcs}} O_{\text{qcs}}^\dagger-O_{\text{qcs}}^\dagger O_{\text{qcs}}\rho_{\text{qcs}}-\rho_{\text{qcs}} O_{\text{qcs}}^\dagger O_{\text{qcs}}\right)$ ($\Gamma_{\text{qcs}}=\left\{\kappa_{d_1},\kappa_{d_2},\kappa_a, \kappa_b, \gamma_{21}, \gamma_{31}\right\}$ and $O_{\text{qcs}}=\left\{d_1,d_2,a, b, \sigma_{12}, \sigma_{13}\right\}$), $\mathcal{L}_{\text{Net},a}\rho_{\text{qcs}}=\sqrt{\kappa_{d_1,\text{ex}_2}\kappa_{\text{ex},1}}([a^\dagger, d_1 \rho_{\text{qcs}}]+[\rho_{\text{qcs}} d_{1}^{\dagger}, a])$, and $\mathcal{L}_{\text{Net},b}\rho_{\text{qcs}}=\sqrt{\kappa_{d_2,\text{ex}_2}\kappa_{\text{ex},3}}([b^\dagger, d_2 \rho_{\text{qcs}}]+[\rho_{\text{qcs}} d_{2}^{\dagger}, b])$, where $\kappa_{d_{1/2}, \text{ex}_i}$ ($i=1, 2, 3, 4$), and $\kappa_{d_{1/2}}$ represent the external and total attenuations of the source cavities, respectively.

We present a full quantum description of the input and output modes by numerically solving Eq.~\eqref{eq:cascadedmaster} in the truncated photon-number state basis of the cavity modes. The input-output relations in the cascaded system are~\cite{PhysRevLett.123.123604.2019,PhysRevA.106.063707.2022}
\begin{equation}
    \begin{aligned}
          & d_{1,\text{out}}  = \sqrt{\kappa_{d_1, \text{ex}_2}}d_1\;, d_{2,\text{out}}  = \sqrt{\kappa_{d_2, \text{ex}_2}}d_2 \;, \\
          & a_t = \sqrt{\kappa_{\text{ex,2}}}a \;,
          a_r = \sqrt{\kappa_{d_1, \text{ex}_2}}d_1 + \sqrt{\kappa_{ex,1}}a \;, \\
         & b_t=\sqrt{\kappa_{\text{ex,4}}}b \;,
          b_r  = \sqrt{\kappa_{d_2, \text{ex}_2}}d_2 + \sqrt{\kappa_{ex,3}}b,
    \end{aligned}
    \label{eq:inputoutputQND}
\end{equation}
with ``$t$'' and ``$r$'' denoting the transmission and reflection coefficients. The operators $d_{1,\text{out}}$ and $d_{2,\text{out}}$ describe photons incoming from the source cavities to the QE-cavity subsystem. Thus, for a signal pulse existing from $t_0$ to $t_1$, the transmittance and reflectance of the cavity modes $a$ and $b$ can be defined as $T_{s}=\int_{t_0}^{t_1}\langle s_{t}^{\dagger} s_t\rangle/\int_{t_0}^{t_1}\langle d_{1,\text{out}}^{\dagger} d_{1,\text{out}}\rangle$ and $R_{s}=\int_{t_0}^{t_1}\langle s_{r}^{\dagger} s_r\rangle/\int_{t_0}^{t_1}\langle d_{1,\text{out}}^{\dagger} d_{1,\text{out}}\rangle$ ($s=a, b$), respectively.

\emph{Quantum Cross Nonlinearity}.---Below we consider symmetric  F-P cavities so that $\kappa_{\text{ex}, 1}=\kappa_{\text{ex}, 2}$ and $\kappa_{\text{ex}, 3}=\kappa_{\text{ex}, 4}$, and $\Delta_a = \Delta_b=0$ for simplicity. Under the bare-cavity approximation without the QE~\cite{SupplMat}, we obtain the mean photon numbers of the two cavities $\langle a^{\dagger}a\rangle\approx2|\alpha|^2/\kappa_a$ and $\langle b^{\dagger}b\rangle\approx2|\beta|^2/\kappa_b$. Here, $|\alpha|^2$ and $|\beta|^2$ are the input photon fluxes, corresponding to the input power~\cite{science.319.1062.2008,PhysRevLett.113.243601.2014,PhysRevA.92.063848.2015}. By solving the steady state of Eq.(\ref{eq:density}), the transmissions for the cavity modes $a$ and $b$ at resonance ($\Delta_1 =\Delta_2 = 0$) are then given by~\cite{SupplMat}
\begin{subequations}
\label{eq:TaTbmain}
    \begin{align}
        T_a &=1 - 8 g_{1}^{2}\left(\kappa_b/\kappa_a\right)\left(\gamma_{21}\kappa_a+2g_{1}^{2}\right) / \mathcal{D} \; , \label{eq:Ta}\\
        T_b &= 1 -  8g_{2}^{2}\left(\kappa_a/\kappa_b\right)\left(\gamma_{31}\kappa_b+2g_{2}^{2}\right)  / \mathcal{D} \;. \label{eq:Tb}
    \end{align}
\end{subequations}
with $\mathcal{D}=\Gamma_A\Gamma_B\kappa_a \kappa_b+16\left( g_{1}^{2}\kappa_b |\alpha|^2+g_{2}^{2}\kappa_a |\beta|^2\right)$, $\Gamma_A = \gamma_{21}+4g_{1}^{2}/\kappa_a$ and $\Gamma_B = \gamma_{31}+4g_{2}^{2}/\kappa_b$.  Note that the term $g_{1}^{2}\kappa_b|\alpha|^{2}+g_{2}^{2}\kappa_a|\beta|^{2}$ in $\mathcal{D}$ appearing in the denominator of Eq.~\eqref{eq:TaTbmain} indicates the crucial dependence of the transmittances $T_a$ and $T_b$ on the excitations of the two cavity modes, despite the \emph{absence} of a direct interaction between them.
The conventional classical cross-Kerr nonlinearity has the form of an ``Ising''-type interaction, $\eta^2\langle a^\dagger a \rangle \langle b^{\dagger}b\rangle$~\cite{PhysRevLett.103.150503.2009,nphotonics.12.744.2018,PhysRevLett.121.203602.2018,PhysRevA.101.053802.2020}. \emph{Remarkably, the quantum nonlinearity in our system has the form $\eta_a\langle a^\dagger a\rangle + \eta_b\langle b^{\dagger}b\rangle$, originating from quantum competition.} Here, $\eta$, $\eta_a$, and $\eta_b$ represent nonlinear coefficients. In this sense, we dub our quantum nonlinearity as ``QCN''.

\begin{figure}
  \centering
  \includegraphics[width=1.0\linewidth]{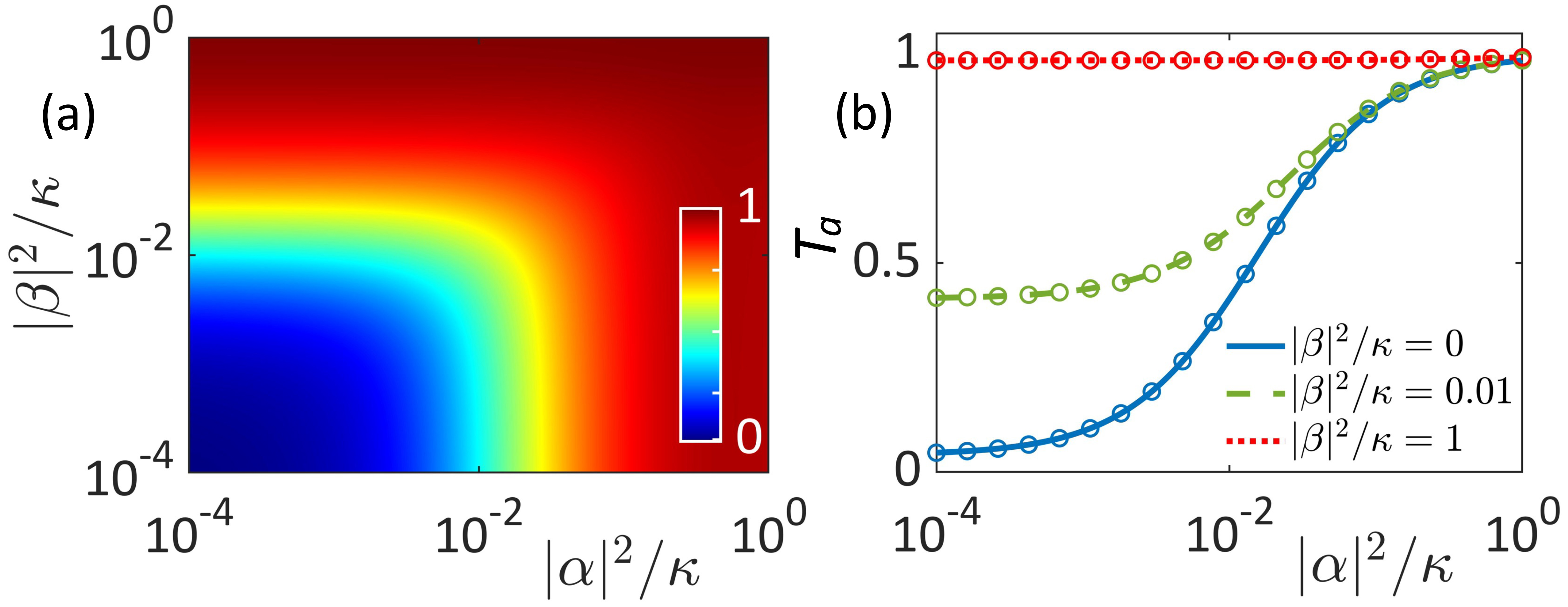} \\
\caption{Quantum cross nonlinearity induced by a V-type QE. (a) Steady-state transmission $T_a$ versus two input powers, $|\alpha|^2/\kappa$ and $|\beta|^2/\kappa$. (b) Transmission $T_a$ as a function of $|\alpha|^2/\kappa$ under different values of
 $|\beta|^2/\kappa$, where curves and open circles are for the analytical and numerical results, respectively. Other parameters: $\kappa_a=\kappa_b=\kappa$, $\gamma_{21}=\gamma_{31}=0.01\kappa$, $g_1=g_2=0.1\kappa$, and $\Delta_1=\Delta_2=\Delta_3=\Delta_4=0$.}
\label{fig:FIG2}
\end{figure}
Without loss of generality, we study the quantum nonlinearity of the cavity mode $a$, characterized by the nonlinear control of transmittance $T_a$ with the input fields, as shown in Fig.~\ref{fig:FIG2}. With identical parameters for the two cavities, the system exhibits a symmetric cross modulation of the transmission $T_a$ in terms of $|\alpha|^2/\kappa$ and $|\beta|^2/\kappa$ [see Fig.~\ref{fig:FIG2}(a)]. Counterintuitively, the transmission $T_a$ can be controlled by the input power $|\beta|^2/\kappa$, even in the absence of direct interaction between the cavity modes $a$ and $b$ [see Fig.~\ref{fig:FIG2}(b)]. This cross modulation implies an application in QND measurement. As $|\beta|^2/\kappa$ increases, the nonlinear curve of $T_a$ displays an overall rise with a decreasing slope with respect to the input power $|\alpha|^2/\kappa$.  Intriguingly, the cavity $a$ can be modified by increasing the driving $\beta$ to become completely transparent to the incident field $\alpha$, even for a weak input power (e.g., $|\alpha|^2/\kappa=10^{-4}$), despite initially being almost completely \emph{absorptive}. This means that the amplitude modulation can be close to $100\%$.

\begin{figure}
  \centering
  \includegraphics[width=1.0\linewidth]{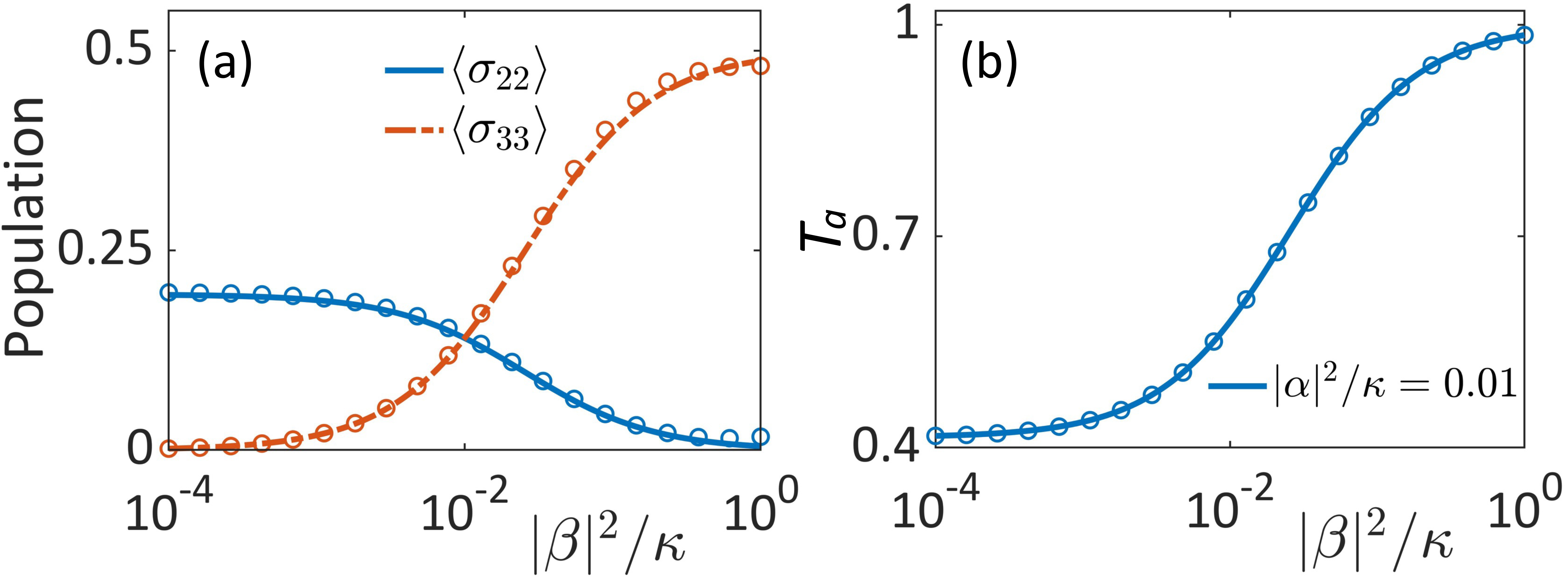} \\
\caption{(a) The population of two excited states of V-type QE, and (b) the transmittance $T_a$ as a function of the input power $|\beta|^2/\kappa$, given a fixed value of $|\alpha|^2/\kappa=10^{-2}$. The analytical results (curves) are in excellent agreement with numerical simulations (dots). Other parameters are the same as those in Fig.~\ref{fig:FIG2}.}
\label{fig:FIG3}
\end{figure}
\emph{Quantum Competition}.---The excitation saturation effect causes the quantum nonlinearity mediated by a two-level QE~\cite{PhysRevLett.113.243601.2014,PhysRevLett.123.233604.2019,nature.569.2019}. In stark contrast, the QCN in our V-type configuration results from an essentially different mechanism---quantum competition between two transitions of the V-type QE, because they share a common ground state. Examining the dependence of the excited-state population on the drivings can provide insights into this striking mechanism. In steady state,  the excited states of the QE are populated to ~\cite{SupplMat}
\begin{subequations}
\label{eq:population}
    \begin{align}
         \langle\sigma_{22}\rangle & = 8g_{1}^{2}\kappa_b|\alpha|^2 / \mathcal{D}  \; , \label{eq:population1}\\
        \langle\sigma_{33}\rangle & = 8g_{2}^{2}\kappa_a|\beta|^2 / \mathcal{D} \; . \label{eq:population2}
    \end{align}
\end{subequations}
According to Eq.~\eqref{eq:population}, the excitations, $\langle\sigma_{22}\rangle$ and $\langle\sigma_{33}\rangle$, compete as a function of the two input powers, $|\alpha|^2/\kappa$ and $|\beta|^2/\kappa$.
Analytical and numerical results are shown in Fig.~\ref{fig:FIG3}(a). For input powers $|\alpha|^2/\kappa=10^{-2}$ and $|\beta|^2/\kappa=0$, the transition $|1\rangle\leftrightarrow|2\rangle$  is dominant, leading to $\langle\sigma_{33}\rangle=0$. However, when $|\beta|^2/\kappa$ increases, {the population is gradually ``drawn" to the state $|3\rangle$ by the $|1\rangle\leftrightarrow|3\rangle$ transition, resulting in a decreasing $\langle\sigma_{22}\rangle$. The two excitations reach a balance when $\alpha=\beta$. As $|\beta|^2/\kappa$ increases further, the $|1\rangle\leftrightarrow|3\rangle$ transition overwhelms the channel $|1\rangle\leftrightarrow|2\rangle$, causing the incident field $\alpha$ to be more likely transmitted compared with the case of the absence of the cavity mode b, see Fig.~\ref{fig:FIG3}(b). This quantum competition leads to the  nonlinear modulation of $T_a$ simultaneously by the input powers $|\alpha|^2/\kappa$ and $|\beta|^2/\kappa$, and in turn changes $T_b$.

QCN is only applicable in the weak-driving regime. At the small photon-number level, the system displays a cross-nonlinear amplitude modulation. However, with a substantially large influx of photons to the cavities, the QE is saturated, and then is prevented from interacting with additional photons. This saturation results in a unitary transmission, as indicated by Eq.~\eqref{eq:TaTbmain}. This QCN promises many novel applications, such as QND measurement of photons.
\begin{figure}
  \centering
  \includegraphics[width=1.0\linewidth]{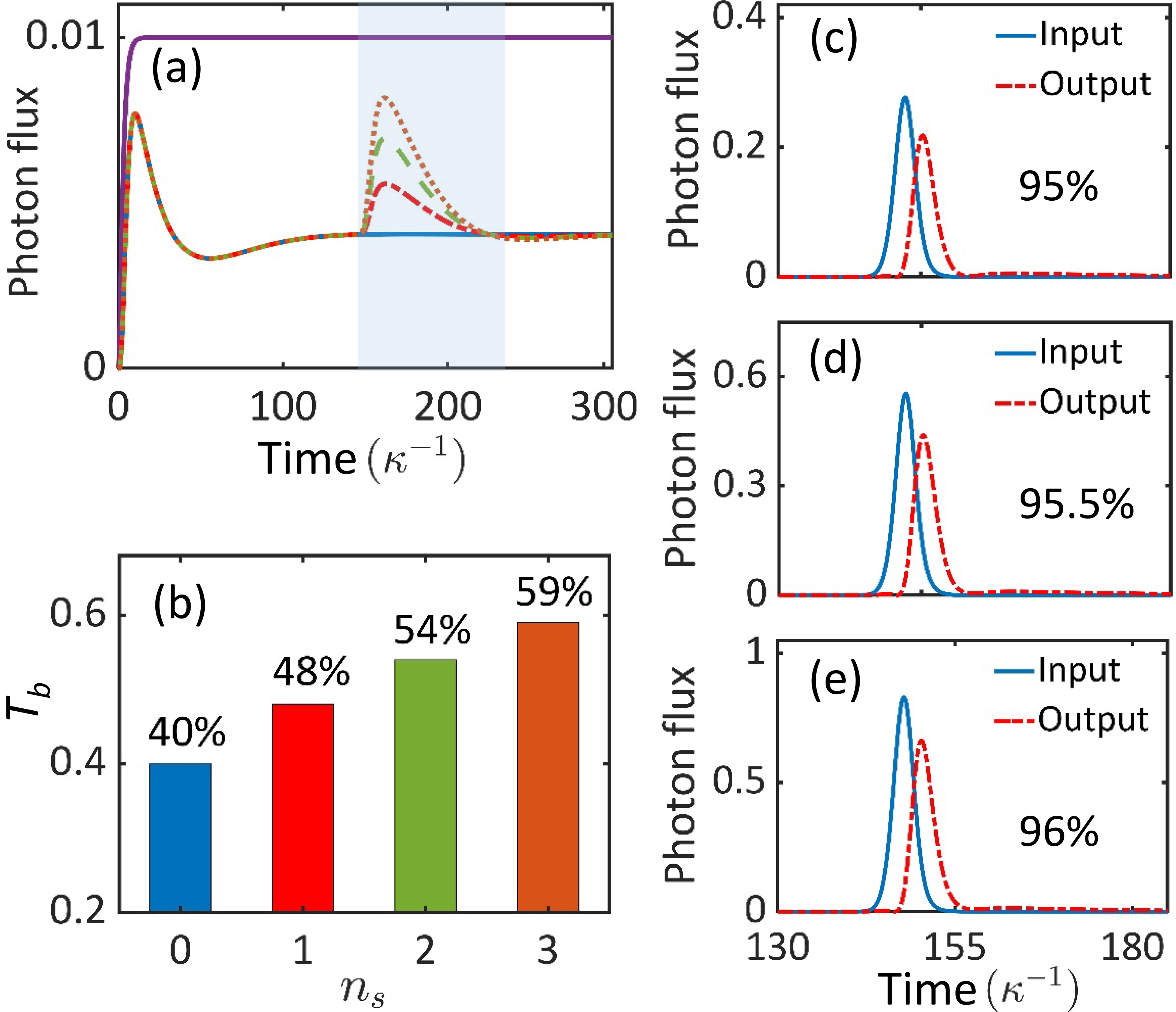} \\
\caption{Photon-number-resolving nondestructive detection based on the QCN. (a) Temporal evolution of the input and output photon fluxes of the probe field. Purple solid curve indicates the probe light incident into the cavity $b$. Blue solid, red dash-dotted, green dashed, and orange dotted curves represent the outgoing photon flux of the probe when the signal pulse contains null, one photon, two photons, and three photons, respectively. (b) Probe transmission vs the signal photon numbers. (c)-(d) Temporal evolution of the input and output probe photon fluxes for different signal photon numbers, with (c) representing the one-photon case, (d) the two-photon case, and (e) the three-photon case. Here, $\kappa_a = \kappa_{\text{ex,1}}=\kappa$, $\kappa_{\text{ex}, 3}=\kappa_{\text{ex}, 4}=0.5\kappa$, and $|\beta|^2/\kappa=10^{-2}$. Other parameters are the same as in Fig.~\ref{fig:FIG2}.}
\label{fig:FIG4}
\end{figure}

\emph{Photon-Number-Resolving Nondestructive Detection}.---Below, we illustrate how QCN can be leveraged for QND measurement. Here, the two input coherent fields, $\alpha$ and $\beta$, are referred to as the signal field and the probe field, respectively. We consider the signal field as a wave packet containing $n_s$ ($n_s=0, 1, 2, 3, \cdots$) photons, and the probe field a continuous weak coherent light. To nondestructively resolve the photon number of a signal pulse, we consider a V-type QE situated in two distinct cavities: a single-sided cavity $a$~\cite{science.342.1349.2013}, featuring one perfectly reflecting mirror and another with a small transmittance for the input and output fields, and a symmetric cavity $b$ in which both mirrors have identical external coupling rates~\cite{SupplMat}, yielding $\kappa_a = \kappa_{\text{ex,1}}=\kappa$, $\kappa_{\text{ex}, 2}=0$, and $\kappa_{\text{ex}, 3}=\kappa_{\text{ex}, 4}=0.5\kappa$. The signal photons are almost entirely reflected off the $a$-mode cavity, while the transmission of the probe field is modulated by the signal photons via the QCN with a photon-number dependence.

The two cavity modes, $a$ and $b$, do not directly interact with each other and separately couple to different transitions of the QE. An intuitive and transparent physical picture of our proposal can be obtained from the steady-state solutions of the system. It can be seen from Eq.~\eqref{eq:Tb} that the transmittance $T_b$ of the probe field increases with increasing the excitation of the cavity mode $a$, namely, the signal photon number. Once signal photons enters the cavity $a$, $T_b$ immediately responds. Consequently, $T_b$ serves as a reliable nondestructive indicator, in real time, for determining the signal photon number.

Using the quantum cascade method~\cite{PhysRevLett.70.2273.1993,PhysRevLett.70.2269.1993,PhysRevA.106.063707.2022,SupplMat}, a full quantum time-domain dynamic evolution simulation is presented in Fig.~\ref{fig:FIG4}. Due to limitation in computational memory resources, we only simulate the system up to a three-photon signal field~\cite{qutip.183.2012,qutip.184.2013}. In the absence of signal photons, namely the null-photon case, the probe light reaches a steady-state transmission approximately at $t=2\pi\times120\kappa^{-1}$ [see the blue solid curve in Fig.~\ref{fig:FIG4}(a)]. To test the modulation of the probe field by a quantum signal, we consider a delayed Gaussian signal pulse, with a delay time $\tau_\text{d} = 2\pi\times150\kappa^{-1}$ and a pulse duration $\tau_\text{s} = 2\pi\times6\kappa^{-1}$. The temporal evolution of the probe transmission is presented in Fig.~\ref{fig:FIG4}(a). The probe transmissions corresponding to different signal photon numbers are indicated by the shadow area. The transmission increases to $48\%$, $54\%$, and $59\%$ when the signal pulse includes one, two and three photons, respectively, see Fig.~\ref{fig:FIG4}(b). This modulation, distinct from the null-photon transmission of $40\%$, clearly confirms the nondestructive detection of the signal photon numbers.

QND measurements require high survival probability of the signal photons~\cite{science.342.1349.2013}. This survival can be evaluated by the reflectance $R_a$ of the signal field. Figures~\ref{fig:FIG4}(c-e) show the input and output signal pulses for $n_s = 1, 2$ and $3$. The corresponding survival probabilities reach $95\%$, $95.5\%$, and $96\%$, respectively, implying a high performance in photon-number-resolving QND measurement. The signal loss attributes to the decay of the state $|2\rangle$ and the intrinsic loss of the cavity mode $a$. Therefore, an effective QND measurement demands the condition $\gamma_{21},\kappa_{\text{in}, a}\ll \kappa_a$~\cite{SupplMat}. Note that our protocol only requires a weak QE-cavity coupling, significantly reducing the difficulty and complexity in experimental implementations.

\emph{Implementation.}---Our QCN requires the two independent transitions of the V-type QE. To meet this requirement, we can separately drive the two transitions with orthogonal polarized cavities~\cite{PhysRevA.90.043802.2014,PhysRevX.5.041036.2015,science.354.1577.2016,PhysRevA.99.043833.2019, science.317.929.2007}. An alternative method involves the use of the different transition frequencies, to avoid mutual interference in excitation~\cite{Natcommun.12.6383.2021}. For a practical implementation, we further assume that the mirrors $\text{M1}-\text{M4}$ are coated with $99.99\%$, $99.2\%$, $99.59\%$, and $99.59\%$ antireflection dielectric layers, respectively. The cavity length of the two miniature F-P cavities is set to $400~\micro\meter$~\cite{nature.436.87.2005}, yielding $\kappa_{\text{ex},1}=2\pi\times480~\mega\hertz$, $\kappa_{\text{ex},2}=2\pi\times6~\mega\hertz$, $\kappa_{\text{ex},3}=\kappa_{\text{ex},4}=2\pi\times243~\mega\hertz$. Considering the impacts of experimental mode matching and other imperfections, we assign a value of $\kappa_{\text{in}, a} = \kappa_{\text{in}, b}=2\pi\times0.5~\mega\hertz$ for the intrinsic loss. We consider the $\text{D}_2$ line of a $^{87}\text{Rb}$ atom for the V-type QE, with levels $|1\rangle=|5^2S_{1/2},F=1,m_{F}=0\rangle$, $|2\rangle=|5^2P_{3/2},F^{\prime}=1,m_{F}^{\prime}=-1\rangle$, and $|3\rangle=|5^2P_{3/2},F^{\prime}=1,m_{F}^{\prime}=1\rangle$. The corresponding two transitions respectively couple to left- and right-circularly polarized cavity modes, with the same wavelength $\lambda=780.2~\nano\meter$. The decay rates of the excited states are $\gamma_{21} = \gamma_{31} = 2\pi\times3~\mega\hertz$~\cite{nature.473.190.2011}. We assume an accessible cavity mode volume of $10^5\lambda^3$~\cite{PhysRevLett.114.023601.2015,PhysRevLett.123.233604.2019,PhysRevLett.130.173601.2023} and then obtain the coupling strengths up to $g_1=g_2=2\pi \times 52~\mega\hertz$. For these experimental parameters, the survival probability can exceed $92\%$.
In addition, by coupling the V-type superconducting qubits~\cite{PhysRevApplied.11.014053.2019,Natcommun.12.6383.2021} with microwave cavities or surface acoustic wave resonators, QCN can be extended to the domains of microwave photons and phonons.

\emph{Conclusions and outlook.}---We have proposed a striking QCN, originating from the unique mechanism of excitation competition between two quantum transition channels. It enables mutual nonlinear control of two quantum bosonic fields. We have further proved the application of this QCN in the number-resolving QND measurement of traveling bosons, with a high probability of survival.

This novel QCN may offer significant applications in quantum information processing, such as quantum logic gates, single-photon switches, and transistors. In the study of condensed matter electronic systems, the quantum competition in population of different ground states has been proposed as a means to understand matter phase transitions~\cite{nature.423.522.2003}, describe quantum criticality~\cite{science.288.475.2000}, and explore novel matter phases~\cite{sciadv.8.35.2022}. Exploring quantum competition in a quantum optical system, as a counterpart, can advance the development of strongly correlated many-body quantum systems, quantum phase transitions, and even time crystals~\cite{natphysYouLi2024}. The revealed cross nonlinear interaction among distinct cavity modes may inspire novel quantum nonlinearity topological physics~\cite{Nat.Phys.20.905.2021, NatPhys.20.1164.2024, PhysRevLett.128.203602.2022} and high-dimensional quantum information processing~\cite{MingyuanPRR.6.033004}.

This work was supported by the Innovation Program for Quantum Science and Technology (Grant No. 2021ZD0301400),  the National Key R\&D Program of China (Grants No.~2019YFA0308700), the National Natural Science Foundation of China (Grants No. 12305020 and No. 92365107), the Program for Innovative Talents and Teams in Jiangsu (Grant No.~JSSCTD202138), the China Postdoctoral Science Foundation (Grant No.~2023M731613), the Jiangsu Funding Program for Excellent Postdoctoral Talent (Grant No.~2023ZB708), and the Natural Science Foundation of Jiangsu Province, Major Project (Grant No.~BK20212004).
F. N. is supported in part by Nippon Telegraph and Telephone Corporation (NTT) Research, the Japan Science and Technology Agency (JST) [via the Quantum Leap Flagship Program (Q-LEAP), and the Moonshot R\&D Grant No. JPMJMS2061], the Asian Office of Aerospace Research and Development (AOARD) (via Grant No. FA2386-20-1-4069), and the Office of Naval Research (ONR) (Grant No. N62909-23-1-2074).
We thank the High Performance Computing Center of Nanjing University for allowing the numerical calculations on its blade cluster system.

 
\providecommand{\noopsort}[1]{}\providecommand{\singleletter}[1]{#1}%

\end{document}